\documentclass{llncs}
\usepackage{url}
\usepackage{graphicx}

\begin{document}

\title{Towards Automated Performance \\Optimization of BPMN Business Processes}
\author{Anastasios Gounaris}
\institute{Dept. of Informatics,
Aristotle University of Thessaloniki, Greece\\
\email{gounaria@csd.auth.gr}
}

\maketitle

\begin{abstract}
Business Process Model and Notation (BPMN) provides a standard for  the design of business processes. It focuses on bridging the gap between the analysis and the technical perspectives, and aims to deliver process automation. The aim of this technical report is to complement this effort by transferring knowledge from the related field of data-centric workflows aiming to provide automated performance optimization of the business process execution. Automated optimization lifts a burden from BPMN designers and increases workflow flexibility and resilience. As a key step towards this goal, the contribution of this work is to provide a methodology to map BPMNv2.0 models to annotated directed acyclic graphs, which emphasize the volume of the tokens exchanged and are amenable to existing automated optimization algorithms.   
In addition, concrete examples of mappings are given, while the optimization opportunities that are opened are explained, thus providing insights into the potential performance benefits and we discuss technical research issues.
\end{abstract}

\section{Introduction}
BPMN has become an international standard for designing workflows. In principle, the basic promise of BPMN is the same diagram prepared by a business analyst to be used for  automating the execution of that process on a modern process engine. This however remains a vision that rarely happens  in practice and the gap between the business and the technical perspectives remains \cite{Stiehl14}. As a result, the executable workflow of business processes is either manually designed in order to provide enterprize-specific configurations or derived by simple procedures using toolkits from established vendors (e.g., Bigazi, IBM, Oracle). Either way, any performance optimization responsibilities rest with experienced IT technicians.

In this work, a different approach is advocated, according to which performance optimization is automated. First, this type of automation relieves  considerable burden from workflow designers. Second, automated optimization yields intrinsically more flexible and resilient workflows. Flexibility and resilience are deemed as both key and particularly challenging aspects in modern BPM \cite{Stiehl14,845FBPM,852BS14}. Increased flexibility stems from the fact that several equivalent alternatives are investigated by the optimizer thus providing more options. Also, when external conditions that impact on the workflow performance evolve, automatically re-optimizing the workflow is important for efficiently adapting to the new setting to attain resilience. Third, performance issues are playing an increasingly important role in modern BPM, which becomes more data- and process-intensive, e.g., in order to cope with big data \cite{Gao13} giving rise to the need for performance optimization. Finally, optimizers for automatically deriving execution details is an integral component in systems that aim to allow end users to submit the process definition at a higher level, e.g., as discussed in \cite{821LNS13}.

Performance optimization is a field that has been largely investigated in databases (e.g., \cite{62Ioa96}) and data-centric flows (e.g., \cite{638VSS+07,691SWCD12,701HPS+12,765OOV+11,KG15,KGfgcs14}). Although these techniques cannot solve the problem of business process optimization in its entirety, they can form a starting for automated performance optimization, as explained in this work. The key first step is to bridge the modeling gap: data-centric flows are typically represented as directed acyclic graphs (DAGs) and optimization techniques rely on statistical metadata, such as cost per task invocation and selectivity, which can be regarded as annotations to these DAGs. We adopt the same modeling abstraction for business processes, and we explain how BPMNv2.0 elements are translated to such annotated DAGs. The intention is to keep using the BPMN standard and the mapping to a DAG, along with the subsequent optimizations, to occur automatically in a way transparent to the process designer.

In summary, the maim contributions of this technical report is  the introduction of an annotated DAG-based approach to BPMNv2.0 modeling along with  concrete examples of mappings of the main BPMNv2.0 elements. A smaller contribution is the presentation of the optimizations enabled with insights into the potential benefits in terms of performance; these optimizations are inspired by their counterparts in the data management area.
For completeness, we  briefly discuss the main technical research issues as well.

In the remainder of this section,  a motivating example and an overview of related work is provided. Next, we elaborate on our DAG modeling abstraction. In Section \ref{sec:bpmn}, the mappings of the main BPMNv2.0 elements to the DAGs proposed hereby are provided. The optimization opportunities and the open research issues are discussed in Sections \ref{sec:opt} and \ref{sec:issues}, respectively. Conclusions are in Section \ref{sec:concl}.

\subsection{Motivating Example}

As a motivating example, we use a sub-process that is encountered in banks for processing loan requests. Upon receiving a customer application, an employee fills in the applicants personal details, and then performs a series of tasks contacting  trusted services from third parties on the web. Such tasks include the following: to import additional customer personal data, to check if the applicant is on any black list, to check the borrowing capacity and to check the information with the help of the national credit bureau. If any of these checks fails, the process aborts and the application is rejected. Finally, the third party services are invoked by providing the customer's SSN identity number.

This scenario is simple but capable of showing a set of optimization issues involved. We give some examples: \emph{Which is the optimal sequence  to contact the third-party services in terms of performance, given that several orderings are valid (e.g., it does not matter whether the check of the borrowing capacity precedes the check regarding the black lists and vice versa)? Should the invocations be performed in parallel? Should an employee fill full personal details only after the checks have passed?} In the envisaged approach, one can take these decisions automatically, in a principled manner in the sense that cost-based algorithms (which may well be accompanied by theoretical optimality guarantees) can be employed. Further discussion on this is deferred to Section \ref{sec:opt}.

\subsection{Related Work}
Automatically devising executable workflows that speed-up execution or improve on other performance metrics is an overlooked area in business process management (BPM).  Optimization in BPM focuses mostly on meeting customer needs and adapting to changing market
conditions \cite{852BS14,Stiehl14}. Performance optimization is considered in the context of \emph{process redesign},  which covers several topics, as discussed in \cite{845FBPM}. Some examples are  to divide an existing process into two or more  separated processes, to eliminate obsolete activities, to assign tasks to the more specialized person and, in general, to perform judicious responsibility assignment, and to buffer requests to external information sources. The most relevant heuristics to database-like optimization are the so-called ``business process behavior heuristics", which include \emph{re-sequencing},  \emph{knock-out}, and \emph{parallelism}.

Re-sequencing covers the optimizations that involve changing the execution order of activities, while preserving the process semantics and correctness. It is the form of optimization that the database community has been investigating since several decades, albeit making different assumptions. A specific form of re-sequencing is to move activities that check conditions, which if not met, lead to process termination, as early as possible. Such activities are termed as knock-out ones. This bears similarities to the filter ordering problem in database queries.
Parallelism deals with decisions as to whether some activities should be executed in a sequential or a time-overlapping fashion. As with re-sequencing, similar decisions are also enabled by database optimizers. We target exactly these form of optimizations, but in a cost-based manner instead of using ad-hoc heuristics.

Also,  there are techniques that restrict their optimizations in the data management tasks within business processes; for example, the proposal in \cite{638VSS+07} tries to consolidate tasks that access databases, thus yielding workflow transformations with fewer and more complex underlying SQL statements, which overall lead to higher  performance. The technique in \cite{668DWS+09} considers BPMN flows but is restricted to optimization of data analytics.
Finally, business processes, similarly to scientific workflows, can be modelled as Petri-nets  or described with the help of $\pi$-calculus and other forms of logic \cite{872CGG09}; such approaches however are very difficult to enable cost-based optimization \cite{765OOV+11} although they allow for other useful functionalities, such as querying as discussed in  \cite{2012Deutsch}.

\section{The proposed DAG-modeled abstraction}
\label{sec:dag}

In data-centric flows, which are also described as DAGs, each graph vertex corresponds to a task. The tasks manipulate data (e.g., extract sentiment information from tweets, combine user identifier numbers with customer info from an underlying database, and so on), and the edges denote how the transformed data flow across the tasks. Since performance in these data-centric flows is directly dependent on the volume of data being processed and the capacity of the execution engines, the optimization methodologies aim to process as fewer data as possible and make judicious assignment of tasks to resources. For the former, the key idea is to prune unnecessary data, that is data that do not contribute to the flow final desired result, as early as possible.

In business processes, the things that flow across tasks are ``tokens". So, our DAGs emphasize the volume of the tokens flowing rather than  the business logic and the control of the flow. Each BPMN task corresponds to a vertex in the DAG. A directed edge connects each ordered pair of vertices, between which a transmission of  tokens takes place in the context of the process.

The goal of the performance optimization is to improve the average performance across multiple process execution.
Performance can be crisply defined in several ways, some of which are discussed in Section \ref{sec:opt}.
Statistical metadata drive the optimization procedure. This metadata are typically extracted from log files. The exact type of metadata depend on the specific optimization problem, but two types are most commonly encountered: \emph{selectivity} and \emph{cost} (per invocation). Selectivity is the average ratio of output to input tokens. For example, a task that, for each given recipe, triggers a task to prepare a single meal has selectivity 1. If a task performs a check and may cause early process termination in case of the test failure, the selectivity is lower than one. Analogously, if a task produces multiple tokens per input token on average, its selectivity is above one. In all the cases,  the output tokens flow across all outgoing edges of the corresponding vertex. The cost of a task is measured in the same units as the performance criterion. If performance is measured in time units, then the task cost is the average time needed to execute that task. In BPMN processes, things (captured by tasks) have to be done under certain circumstances (captured by gateways); in our DAG-based approach, the statistical metadata play an important role in considering the gateways semantics through annotations to vertices.

Apart from the statistical metadata, there are two other categories of metadata needed. The first category covers dependency constraints between task pairs, i.e., whether a task must precede another one in any execution plan to preserve semantic correctness or whether two tasks must be placed on distinct DAG branches, where each DAG branch corresponds to a different execution path. The second category captures behavioral characteristics, such as whether the task can be parallelized, so that its workload is executed by multiple executors in parallel, and whether a task can operate in a pipelined manner, i.e., to be capable of producing output before consuming its entire input.

A distinctive feature of our proposal is that BPMN tasks correspond to DAG vertices but the opposite does not necessarily hold. We employ the notion of artificial tasks termed as \emph{dummy} tasks. Overall, the combination of normal and dummy tasks with appropriately set statistical metadata allow for modeling the token flow in business processes and paves the way for performance optimization, as discussed next.

\section{BPMNv2.0 Symbol Mapping}
\label{sec:bpmn}

In this section, we describe how we can model the main elements of BPMNv2.0 to our annotated DAGs. The examples are deliberately simple, to convey easier our message, and they are drawn from the Camunda platform.\footnote{\url{http://camunda.org/bpmn/reference/}} To avoid confusion, the tasks in our DAGs are depicted as circles rather than rounded rectangles, following also the standard convention for vertices in graphs.

\subsection{Activities}

\subsubsection{Task.}

\begin{figure}[b!]
 \centering
 \includegraphics[width=0.8\textwidth]{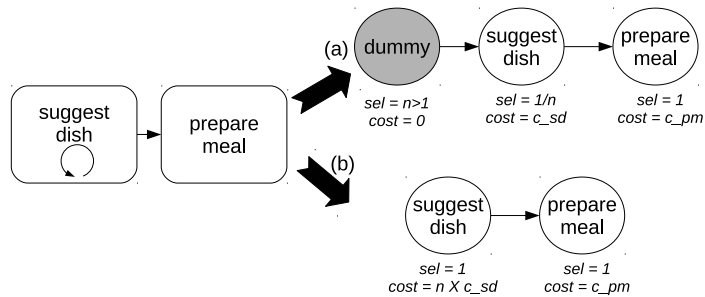}
\caption{Mapping a  loop and an ordinary task.}
\label{fig:loop1}
\end{figure}

An ordinary task, independently of its exact type (e.g., manual, service, business rule, and so on), is mapped to a distinct vertex in our model. The cost metadata is the average cost in time units to execute that task, and its selectivity is the average ratio between the output and input tokens.

For example, in Figure \ref{fig:loop1}, there is the task \emph{``prepare meal"}. This task is triggered after it has received the menu suggestions from the previous task. If the average time to prepare a meal is \emph{c\_pm}, then the cost of that vertex takes that value. The selectivity is set to 1, because for each suggestion there is a single meal prepared.

The loop tasks, like the \emph{``suggest dish"} one in the figure, require a bit more attention, because they execute more than one time on average. Let us suppose that the average number of times the task is activated is $n$ and the average time cost to execute each time is  \emph{c\_sd}.
There are two ways to handle this case. First, we can insert a zero cost \emph{dummy} task before \emph{``suggest dish"}.  The selectivity of the dummy task is set to $n$, whereas the cost of the vertex corresponding to \emph{``suggest dish"} remains  \emph{c\_sd}. However, the selectivity of this vertex needs to become $1/n$ to account for the fact that even if $n$ times the \emph{``suggest dish"}  task is executed, there is always one token passed on to the subsequent task. The second option is not to use a dummy task and amortize the cost of the vertex, so that it captures the fact that, on average, it is executed $n$ times and thus becomes $n \times c\_sd$.

Both options are shown in Figure \ref{fig:loop1}.  The second one is simpler, but it hides the information about the average number of iterations. In cases where this is required, e.g., to devise optimized plans or compute metrics that are parameterized according to $n$, then the first approach should be preferred.

In the previous examples both tasks are not parallelizable. The multiple instance tasks can be modeled in the same way as the loop ones, but the difference is that they can be parallelized up to a parallelism degree of $n$.

\subsubsection{Compensation Tasks}

\begin{figure}[tb!]
 \centering
 \includegraphics[width=0.8\textwidth]{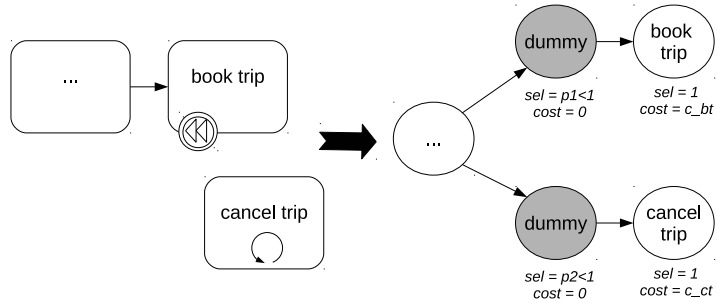}
\caption{Mapping a compensation task.}
\label{fig:compensation1}
\end{figure}

Compensation tasks can be mapped to our DAG with the help of dummy tasks as well. Consider the example in Figure \ref{fig:compensation1}, where a \emph{book trip} task with cost \emph{c\_br} is associated with a compensating task \emph{cancel trip} with cost \emph{c\_ct}. Let as also assume that the probability of not triggering the compensating task and continuing the normal execution is \emph{p1}, whereas the probability of canceling the trip is \emph{p2=1-p1}. The mapping to our DAG involves two dummy tasks, which do not contribute to the cost, but control the amount of flow to each of the two branches in a way proportional to the afore-mentioned probabilities through setting their selectivities accordingly. The preceding task in this example sends its output to both branches in line with the edge interpretation in our DAG model, and it is the responsibility of the dummy tasks to perform the filtering. The dependency constraints state that none of the two initial tasks should precede the other, i.e., the optimizer cannot place them in a sequence.

Note that a simpler mapping would also be possible in cases where the two branches merge just after the book and cancel tasks. In that mapping, we could omit the dummy tasks and have only the  \emph{book trip} task with selectivity set to 1 and a weighted cost equal to $p1 \times c\_bk + p2 \times c\_ct$. This mapping does not capture the complete business logic, but is adequate for performance optimization. Similarly, if there are subsequent tasks following \emph{book trip} but no output edge for \emph{cancel trip}, we could have a single task with the weighted cost as above and the selectivity being equal to $p1$.

\subsubsection{Subprocess.} Subprocesses do not pose any specific challenge per se with regards to their mapping. However, for optimization purposes, it is always more desirable to expand them in order to broaden the optimization search space of the algorithms, provided that those algorithms are capable of navigating efficiently through the expanded search space.

\subsubsection{Call Activity.} From the performance point of view, call activities can be treated like ordinary activities in the way described above.

\begin{figure}[tb!]
 \centering
 \includegraphics[width=0.8\textwidth]{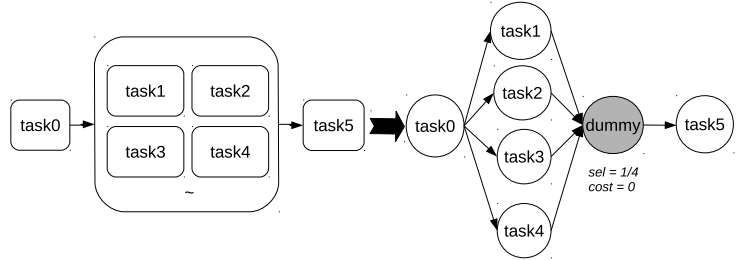}
\caption{Mapping an ad-hoc task.}
\label{fig:adhoc1}
\end{figure}
\subsubsection{Adhoc.} Adhoc subprocesses contain several tasks that can be executed at any order. This is exactly the sweet spot for database-like optimization, which can decide on the optimal order in a principled manner. In Figure \ref{fig:adhoc1}, we present an example with an adhoc subprocess with 4 tasks that can be executed in arbitrary order. We map them in the way shown in the right part of the figure; all the tasks in the adhoc subprocess are directly connected to the preceding task to denote that there are no inter-dependencies among them and, as such, can be computed in parallel (although the final decision rests with the optimizer as discussed later). Also, we use a zero-cost dummy combiner task to aggregate the output tokens of the adhoc tasks and call the next activity. The selectivity of that dummy task is set to 0.25 because it outputs one token for every four tokens received as input. Further, it is not pipelining, because it needs to consume all its four input tokens in each execution, before creating an output token.

\begin{figure}[tb!]
 \centering
 \includegraphics[width=0.8\textwidth]{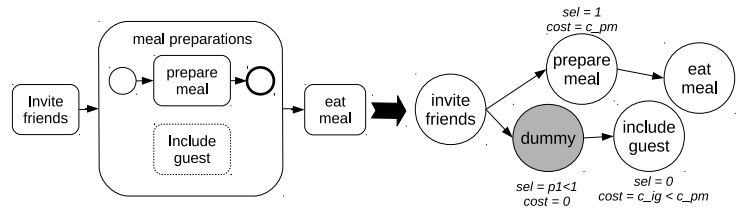}
\caption{Mapping an event subprocess.}
\label{fig:eventsubpr1}
\end{figure}

\subsubsection{Event Subprocess.}

An event subprocess may be executed while the enclosing subprocess is active. An example is presented in Figure \ref{fig:eventsubpr1}, where the enclosing process is a task  \emph{prepare meal}, during which new guests can be included (captured by the event task \emph{include guest}). The costs of these tasks are \emph{c\_pm} and \emph{c\_ig}, respectively.
To map this case to our token-flow DAG, we insert a dummy filtering vertex with selectivity equal to the probability of executing the event subprocess \emph{p1}. Note that, by definition,  \emph{c\_ig} is always less than \emph{c\_pm}, and the two activities cannot be executed sequentially.

\begin{figure}[tb!]
 \centering
 \includegraphics[width=0.8\textwidth]{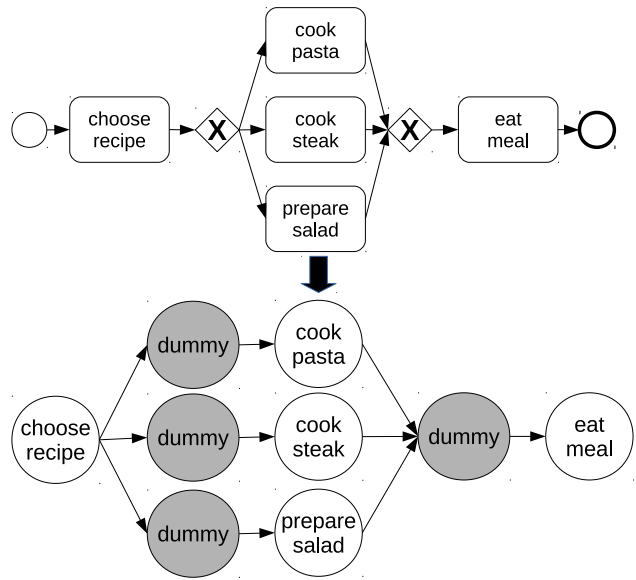}
\caption{Mapping an exclusive gateway.}
\label{fig:exclusive1}
\end{figure}
\subsection{Gateways}

\subsubsection{Exclusive, Parallel and Inclusive.}

Gateways is a core BPMNv2.0 element and a distinctive feature of process-centric flows not appearing in data-centric workflows. As such, their effective mapping is of high significance in our approach.
We distinguish between exclusive, parallel and inclusive gateways. An example of an exclusive gateway is in Figure \ref{fig:exclusive1}, where there is an option during meal preparation for the three dishes shown. Let us assume that the average probability to select each of the three options is \emph{p1}, \emph{p2}, and \emph{p3}, respectively;  these probabilities need to sum to 1 to account for the fact that always exactly one option is selected. Then, we can insert zero cost dummy filtering tasks with their selectivity set equal to the probabilities above. There is also a zero-cost unary-selectivity dummy combiner task being responsible for synchronization, but this is optional. To preserve the gateway semantics, the dependencies between these vertices are that they cannot be placed in a sequence.

Now, suppose that, in the previous example, the gateway is transformed to a parallel one. Then the dummy tasks on the left become optional (corresponding to zero cost and selectivity equal to 1), so that the three previous options
can be directly connected to \emph{choose recipe}. However, the dummy task on the right becomes compulsory and its selectivity is set to 1/3, derived by a generic formula: ratio of 1 to the number of tasks after the parallel gateway. This is to ensure that \emph{eat meal} receives a single token for each \emph{choose recipe} execution no matter how many intermediate tasks are executed in parallel. In addition, the dummy task on the right enforces synchronization.

Inclusive gateways allow tokens to flow across one, many or all paths, i.e., the combine certain features from the exclusive and parallel cases. The high-level mapping  shown in Figure \ref{fig:exclusive1} holds for inclusive gateways as well, but with all dummy tasks being compulsory. Contrary to the case of exclusive gateways, the selectivities of the dummy tasks preceding the options can sum to a value greater than 1.  In addition, the dummy combiner task on the right part becomes compulsory (as in parallel gateways) and its selectivity is set to the ratio of 1 to the sum of the selectivities on the left.

\begin{figure}[tb!]
 \centering
 \includegraphics[width=0.8\textwidth]{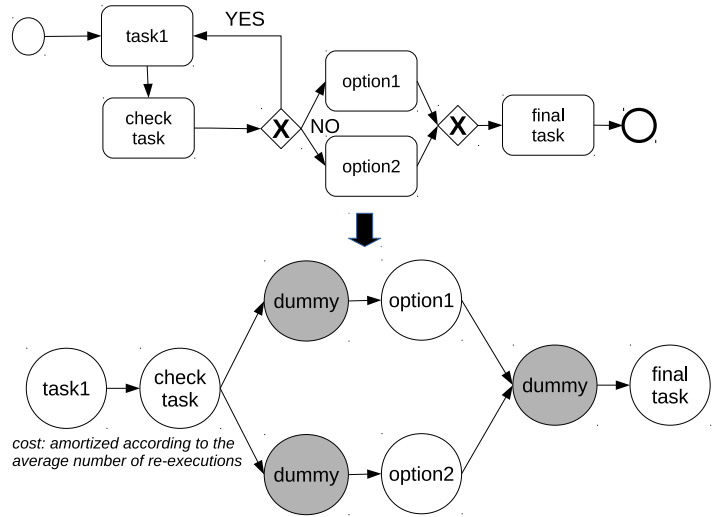}
\caption{Mapping a  more complex exclusive gateway.}
\label{fig:loop2excl}
\end{figure}
\subsubsection{Gateways and Loops.} Gateways are very common to be accompanied by cycles in business graph models, as shown at the top of Figure \ref{fig:loop2excl}. We can combine the approaches presented earlier regarding loops and gateways in order to render our graph acyclic. The tasks belonging to the loop path are placed in a sequence with their cost having been amortized  as shown in Figure \ref{fig:loop1}; the alternative of a dummy task in that figure is also valid.
Then, the rest of the tasks after the gateway are treated as in Figure \ref{fig:exclusive1}.

\subsubsection{Event-based vs. Data-based Gateways}

In BPMNv2.0, there is a distinction between data-based and event-based gateways. The former choose the routing of a token to one or more paths according to data associated with the specific token. The latter make decisions based on events happening. From the performance point of view, there is no essential difference between these two cases. For example, consider the exclusive case. Instead of monitoring the probabilities of task activation, in an event-based exclusive gateway, we can monitor the probability of corresponding events and set the selectivities in our DAG vertices accordingly. Next, we discuss BPMNv2.0 events in detail.

\subsection{Events}

In general, events impact on the task statistical metadata from the time performance perspective. Not all events need to be mapped to our DAG though. For example, process starting need not be explicitly defined, since we are interested in modeling and optimizing the average performance of process execution rather than on enacting the processes.

\begin{figure}[tb!]
 \centering
 \includegraphics[width=0.8\textwidth]{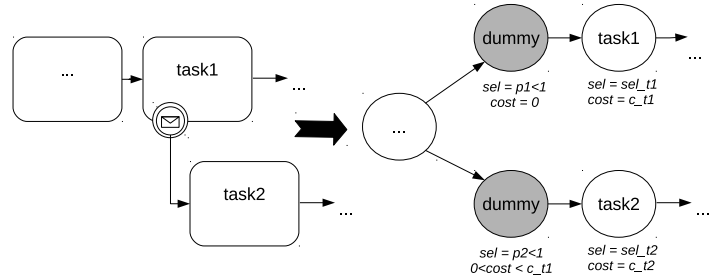}
\caption{Mapping a  subprocess with a boundary  event.}
\label{fig:boundary-event1}
\end{figure}

\subsubsection{Boundary Events.} Boundary events are intermediate events that may cause  interruption of task execution. Consider the example in Figure \ref{fig:boundary-event1}. \emph{Task1} starts its execution and, at some
while it is still active, it may be interrupted if a boundary event occurs. The interruptions causes immediate termination of \emph{task1} and the token is passed to \emph{task2}.  If no event happens, \emph{task1} is completed normally and the token is passed to its next task.

Let us suppose that the costs (resp. selectivities) of the two tasks are \emph{c\_t1} and \emph{c\_t2} (resp. \emph{sel\_t1} and \emph{sel\_t2}). Also, the probability of \emph{task1} executing normally is \emph{p1} and the probability of an event is \emph{p2=1-p1}. In our DAG, we create two branches to denote each execution mode. At the beginning of each branch, we place a dummy filtering task with the selectivity equal to the probability of following the corresponding path. There is one more subtle detail. When \emph{task1} is interrupted, it has already incurred some cost, which needs to be captured. Let us suppose that, on average, when the interruption takes place, this is after \emph{cost} time units; this quantity becomes the cost metadata for the dummy vertex of the event-related branch.

BPMNv2.0 also accounts for non-interrupting boundary events. In the previous example, if the event was non-interrupting, the execution of \emph{task1} would always complete. In that case, the difference in the mapping will be that the dummy activity on the normal path is placed \emph{after} the task. The rest remains the same.

\subsubsection{Message.} Messages play an important role in BPM but are not related to performance. As such, they can be omitted. In case we want to capture them explicitly, e.g., as if they play a checkpointing role, we can insert corresponding dummy vertices with zero cost and unary selectivity.

\begin{figure}[tb!]
 \centering
 \includegraphics[width=0.55\textwidth]{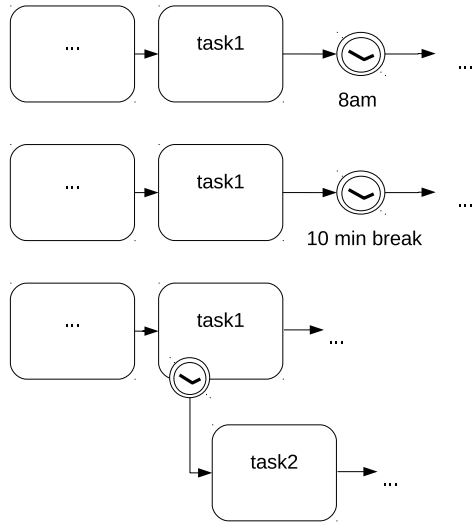}
\caption{Three cases of using timer  events in BPMNv2.0.}
\label{fig:timer-event1}
\end{figure}

\subsubsection{Timer.} Timer events are important for performance in time units, and need to be handled with care according to their semantics. We examine three representative cases of timer event usage, as shown in Figure \ref{fig:timer-event1}. In the first case, we execute a subprocess and the second subprocess starts its execution at 8am. From the performance optimization point of view, this event inserts a time barrier and implies that the two subprocesses need to be optimized for performance independently. That is, we need to create two DAGs, one for the subprocess before the timer event, and one for the subprocess after it.

The second case is different and refers to a time break of 10 minutes. In that case, we insert a dummy task with unary selectivity and time cost equal to the amount of time units, which is equivalent to 10 minutes. Third, time events can play the role of boundary events considered above.  One important difference however is that, if these events are interrupting, the time cost of \emph{task1} should not be its average cost in time units, but the average cost when the event is not triggered. Analogously, the cost of the dummy activity on the event path should be equal to the amount of time passed in order to trigger the event.

\subsubsection{Error and Compensation.} In general, errors may not be considered, given that performance optimization makes more sense when everything is executed correctly. For the specific cases where errors trigger subprocesses and compensation tasks, the methodologies described above apply. The same holds for the compensation events, too.

\subsubsection{Conditional.} Conditional events may postpone the token propagation to the subsequent tasks until a set of conditions is satisfied. This affects the time performance, and to capture the impact we need to insert a dummy vertex at the place of the conditional event with cost proportional to the amount of time waiting in order to meet the requirements. In general, the selectivity is set to 1; if there are execution examples, where the conditions are never met so that the events acts also as a filter, then the selectivity is configured accordingly.

\subsubsection{Signal.} Signal events have no differences with messages from the performance point of view.

\subsubsection{Termination.} This type of events affects the amortized cost of all processes that may be terminated. For example, if a task has cost \emph{c1} in \emph{p1\%} of the cases, and in the rest, it is terminated due to a terminate event after \emph{c2} time units, its amortized cost becomes $p1 \times c1 + (1-p1) \times c2$.

\subsubsection{Cancel.} Cancel events affect the selectivity of subsequent transaction processes.

\subsubsection{Other Event Types.} BPMNv2.0 provides for further event types, which however, do not require special treatment. Such types include the links, multiple and parallel, and escalation.
Multiple and parallel provides insights into how often a complete subprocess is triggered, thus possibly affecting the selectivities of ordinary and dummy tasks.
Finally, the behavior of escalation from the performance point of view is covered by elements such as gateways.


\section{Optimization Opportunities}
\label{sec:opt}

\begin{figure}[tb!]
 \centering
 \includegraphics[width=0.8\textwidth]{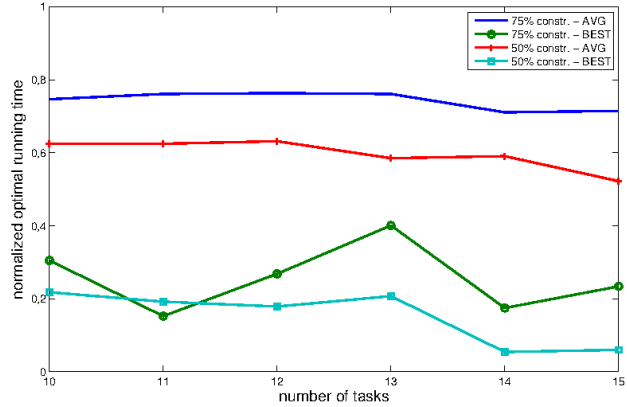}
\caption{Example benefits when optimally reordering activities.}
\label{fig:ex2}
\end{figure}

The optimizer envisaged, can take the initial mapping, and the set of statistical and dependency metadata and derive an optimal execution plan. Two main optimization approaches are to decide  the exact order of task execution and the assignment of tasks to executors (either engines or humans) in a cost-based manner.

To provide insights into the benefits, we extend the motivating example, where the ordering of some tasks is flexible thus generalizing adhoc tasks in BPMN. Suppose a simple process, which contains $n$ activities forming a chain (i.e., there are no branches).  If, due to dependency constraints, there is no flexibility in the order, then this implies that there are $\frac{n(n-1)}{2}$ dependency edges, either explicitly stated or implied through transitive closure; e.g., if task \emph{A} must precede \emph{B}, which must precede \emph{C}, then it is inferred that \emph{A}  must also precede \emph{C}. In practice, this rarely happens. For example, a loan pre-processing template may define the order in which the tasks for importing contact information of the applicant, checking the borrowing capacity and contacting the credit bureau take place, despite the fact that any ordering is valid.

Figure \ref{fig:ex2} shows the performance improvements over 100 randomly generated DAGs, where there are $0.75\frac{n(n-1)}{2}$ and $0.5\frac{n(n-1)}{2}$ dependency edges, $n$ ranges from 10 to 15, the selectivity of tasks ranges from 0.01 (extremely filtering) to 2, and the cost ranges from  1 to 100 (i.e., the cost of the most expensive task is two orders of magnitude higher than the cost of the less expensive one). In all DAGs, the exact values were randomly chosen with uniform distribution. The exact optimization metric selected is the  sum the average execution time  for each task; the latter is defined as the product of the task cost and selectivity values of the preceding tasks. This optimization metric defines the average running time of the process. As baseline performance, which corresponds to normalized value 1, we consider the running time of the initial DAG before re-orderings. In the figure, we can see that, on average, there is a reduction in the running time by 25.62\% for the more constrained case; moreover, the average reduction becomes 40.37\%, for 50\% constraints. Also, there are isolated runs, where the improvements can be of several factors (up to an order of magnitude), as shown by the maximum improvement plots at the bottom of the figure. These numbers indicate how significant the performance benefits can be, even in simple processes.

Of course, exploring all the orderings, even in highly constrained settings, is an intractable problem. The techniques in \cite{KG14} show how the optimizer can navigate through the search space  in a scalable manner.
Other performance optimization problems can be considered as well. For example, the technique in \cite{421wsms06} is applicable to a variant of the previous setting, where all tasks are executed in parallel and some of these tasks are acting as performance bottlenecks, e.g., because they are manual ones; in that case, the optimization goal is to order the tasks in such a way, so that the average bottleneck is mitigated as much as possible. A nice feature of this technique is that it automatically decides whether a task should be executed in parallel with others or not. Another flavor of performance problems is to minimize the sum of the task costs across the critical execution path of the business process DAG rather than considering the complete process, leveraging the techniques in  \cite{734ABDR12}. As a final example of optimization goals enabled by task re-ordering, \cite{721DH12} presents techniques that maximize the throughput (i.e., equivalent to token consumption rate) through maximizing the utilization of each executor.

In addition and again with a view to improving performance, task assignment can be performed in a principled manner. As an example, we mention the approach in \cite{KGfgcs14}, where the different execution options for each task are checked in a scalable way, while taking into account realistic aspects, such as the overhead when switching between executors (e.g., two adjacent manual tasks are delegated to different persons) and that each task can be executed only by a restricted set of executors (e.g., a manual task cannot be performed by any employee).




\section{Main Research Issues}
\label{sec:issues}

Here, we mention the main research issues for developing complete solutions for performance optimization in BPMN business processes.

\begin{itemize}

\item \emph{Need for dependency-aware optimization algorithms.} Mapping BPMN models to our DAG abstraction is a necessary but not sufficient condition to perform cost-based performance optimization. In the previous section, we referred to several algorithmic techniques; albeit those techniques cannot apply to generic DAGs produced by the proposed mapping from BPMN. The reason is that, those algorithms consider only precedence constraints, i.e., constraints of the form that task \emph{A} must precede task  \emph{B}. This type of constraints needs to be complemented by (i) parallelism constraints that enforce tasks to be placed in different execution paths; (ii) blocking vs. pipelining information for each task; and (iii) parallelism capability information, to define which tasks are amenable to parallel execution and up to which degree of parallelism. Consequently, more research is needed in optimization algorithms to develop solutions that account for the complete range of constraints in business processes.

\item \emph{Statistical Metadata Collection.} The statistical metadata play a crucial role, and their efficient collection requires special attention.  Techniques like the one in \cite{823HDP14} and histograms \cite{62Ioa96} may act as a starting point. Challenges include the fact that statistics are actually correlated, which means that changing the order of tasks may affect their statistical metadata. For example, a timer boundary event may fire less frequently, if the optimizer assigns the corresponding task to a more powerful execution engine, so that it takes less time to complete on average.
 This again relates to the need for more advanced optimization algorithms that can handle such correlations of statistical values.

\item \emph{Extensive Evaluation.} There needs to be extensive evaluation using benchmarks to reason with confidence about the actual capability of each proposed optimization technique. TPC\footnote{\url{http://www.tpc.org/}} has developed such benchmarks for database queries but no related efforts for business processes exist to date.

\item \emph{Mapping to BPMN models and end-to-end solutions.} In this work, we showed how we can map BPMN models to our DAGs and perform optimizations under a specific case of dependency constraints. Holistic solutions should involve the mapping of the optimized execution plan back to a BPMN model. In general, there is a need for end-to-end solutions that ideally would be exposed as a software plugin to existing platforms and encapsulate all the mapping and optimization methodologies in order to render the optimization fully transparent to the process designer.

\end{itemize}

Finally, it should be remarked that, in BPM, optimization for performance only is inadequate; after addressing the issues above, the focus should also be shifted to aspects such as fault-tolerance  \cite{632SWCD09}, reliability, economic cost and so on, potentially building on top of multi-objective query optimization \cite{329PY01,TGT2013}. This needs also to be complemented by theoretical investigation, since there is no theory of declarative business process optimization that is comparable to the one in databases \cite{2012Deutsch}.

\section{Conclusions}
\label{sec:concl}

This work is motivated by the fact that currently, performance optimization of business process is a manual activity in the responsibility of the designer. Due to the complexity of modern processes and/or the volatility of the execution environment, there is no performance optimality guarantee. To address this limitation, automated performance optimizations should be applied. We explain how we can build upon the knowledge in the data management community to optimize data-intensive queries and flows. More specifically, we discuss the annotated DAG modeling abstraction required to employ such solutions, going through the handling of the main BPMNv2.0 elements in detail. Using a concrete example, we provided insights into the potential performance benefits, which can be of an order of magnitude. Finally, we identified the main research issues for enabling automated optimization in BPM.

\bibliographystyle{splncs03}
\bibliography{../bpm}

\end{document}